\documentstyle[preprint,revtex]{aps}
\begin{document}
\draft
\preprint{IP/BBSR/91-32}
\begin{title}
Hot nuclear matter : A variational approach
\end{title}
\author{H. Mishra, S.P. Misra, P.K. Panda}
\begin{instit}
Institute of Physics, Bhubaneswar-751005, India
\end{instit}
\author{B.K. Parida}
\begin{instit}
Physics Department, Regional College of Education,\\
Bhubaneswar-751007, India
\end{instit}
\begin{abstract}
We develop a nonperturbative technique in field theory to study
properties of infinite nuclear matter at zero temperature as well as
at finite temperatures. Here we dress the nuclear matter with off-mass shell
pions. The techniques of thermofield dynamics are used for finite temperature
calculations. Equation of state is derived from the dynamics of the
interacting system in a self consistent manner.
The transition temperature for nuclear matter appears to be around 15 MeV.
\hfil\break

\vspace {1in}
\noindent (To appear in Int. J. Mod. Phys. E)
\end{abstract}

\section {Introduction}

The understanding of hot dense nuclear matter is an interesting problem
in theoretical physics in the context of heavy ion collision experiments
as well as big bang cosmology. The problem here is basically nonperturbative.
The interaction of nucleons which may arise as a residual interaction
due to their substructure of quarks and gluons is technically
not solvable. The usual approach
is to tackle the problem through meson nucleon interactions. This also
entails a nontrivial technical problem with the pion nucleon coupling
$G_{NN\pi}^2/4\pi$ being as large as 14.6 making any perturbative calculation
unreliable. An alternative consistent theoretical frame work
has been developed by Walecka$^1$ consisting of
interactions of nucleons with a neutral scalar field $\sigma$ as well as a
neutral vector meson $\omega$. This has been
done at zero temperature as well as finite temperatures.$^2$
Variations of the
same model has also been considered including cubic and quartic
terms in the ${\sigma}$ fields to reproduce correct bulk modulus of
nuclear matter.$^3$
These calculations however use meson fields as classical, and, use a
$\sigma $-field which is not observed.

An alternative model for infinite nuclear matter consisting of interacting
nucleons and pions was considered in Ref [4]. The nuclear matter
was dressed with off mass shell pion quanta. The scalar
isoscalar pion condensates
simulated the effects of $\sigma$ mesons$^5$ with the short distance repulsion
arising from composite structure of nucleons and/or through vector meson
exchanges. This is not only aesthetically appealing with clasical $\sigma $
fields arising from quantum mechanical structures but also has a stronger
phenomenological appeal as $\sigma $ mesons have not been found in nature.
With a similar approach we shall consider nuclear matter at
finite temperatures. The methods of thermofield dynamics$^6$
fit naturally for this purpose because
here statistical average is done through an expectation value over a
"thermal vacuum" in an extended Hilbert space.

The outline of the paper is as follows.  In Sec. II we shall consider
nuclear matter with pion condensates at zero temperature. Here we shall
briefly summarize the results of Ref. [4] and show that these results
constitute a particular approximation
of the present method.
In Sec. III we shall discuss
finite temperature nuclear matter using the methods of thermofield
dynamics.$^6$
We shall calculate thermodynamic quantities like pressure, entropy
density and energy density for nuclear matter and derive the
equation of state for the same. Section IV will consist of discussions
of the results obtained in the present model. A possible experimental signature
resulting from the model is discussed in Sec. V. In Appendix A
we shall summarise some results of thermofield dynamics$^6$ for the sake
of completeness.

\section {Zero temperature formalism}

We shall here start with the effective Hamiltonian for pion nucleon
interactions which was derived in Ref. [4] and given as
\begin{equation}
{\cal H}_N(\vec x)={\cal H}_N^{(0)}(\vec x)+{\cal H}_{int}(\vec x)
\end{equation}
where the free nucleon part ${\cal H}_N^{(0)} $ is given as
\begin{equation}
{\cal H}^{(0)}_N(\vec x)=\psi_I^\dagger(\vec x) \epsilon_x \psi_I(\vec x)
\end{equation}
and the effective pion nucleon interaction part is given as
\begin{equation}
{\cal H}_{int}(\vec x)=\psi_I^\dagger(\vec x) \left[
-{iG\over 2 \epsilon_x }\vec \sigma.
\vec p \phi +{G^2\over 2 \epsilon_x }\phi^2\right]\psi_I(\vec x).
\end{equation}
In the above, $\epsilon _x=(M^2-\vec\bigtriangledown_x^2)^{1/2}$ where
$M$ denotes the nucleon mass.
Furthermore the free meson part of the Hamiltonian is
given as
\begin{equation}
{\cal H}_M(\vec x)={1\over 2}\left[{\dot \phi}_i^2
+(\vec \bigtriangledown \phi_i)\cdot(\vec \bigtriangledown \phi_i)
+m^2\phi_i^2\right]
\end{equation}
Clearly in the above $\psi_I$ refers to the large component of the
nucleon spinor and $\phi=\tau _i \phi_i$.

We expand the field operator $\phi_i(\vec z)$ in terms
of the creation and annihilation operators of off-mass shell mesons
satisfying equal time algebra as
\begin{equation}
\phi_i(\vec z)={1\over \sqrt{2 \omega _z}}(a_i(\vec z)^\dagger +a_i(\vec z))
\end{equation}
and
\begin{equation}
\dot\phi_i(\vec z)=i{\sqrt{\omega _z\over 2}}(a_i(\vec z)^\dagger -a_i(\vec z))
\end{equation}
where we take with the perturbative basis $\omega _z=(m^2-\vec \bigtriangledown
^2_z)^{1/2}$, with $m$ denoting
the mass of the meson.

The two pions in Eq.(3) constitute a scalar-isoscalar
interaction of nucleons and thus could simulate the
effects of $\sigma $-mesons.
With this in mind, let us consider a two pion creation operator
given as
\begin{equation}
B^\dagger ={1\over 2}\int \tilde f(\vec k)a_i(\vec k)^\dagger a_i
(-\vec k)^\dagger d \vec k
\end{equation}
where the arbitrary function $\tilde f(\vec k)$ will be determined
by a variational procedure.
We now introduce a 'meson' dressing of nuclear matter through
the state
\begin{equation}
|f>=U|vac>
\end{equation}
where
\begin{equation}
U=exp(B^\dagger-B)
\end{equation}
In contrast to Ref . [4], the operator $U$ is unitary, and thus we have
\begin{equation}
<f| f>=1
\end{equation}
This makes the formalism simpler and more realistic.
Further with this ansatz, one easily obtains
\begin{equation}
U^\dagger a_i(\vec k)U=(cosh\tilde f(\vec k))a_i(\vec k)+
(sinh\tilde f(\vec k))a_i(-\vec k)^\dagger
\end{equation}
which is a Bogoliubov transformation to be used later for the calculations.

We shall take $N$ nucleons occupying
a spherical volume $V=4\pi{R^3\over 3}$, where $N/(4\pi{R^3\over 3})=\rho $
remains constant as $N\rightarrow\infty$ and we neglect the surface effects. We
describe the system with the density operator $\hat \rho _N$ such that$^4$
\begin{equation}
tr\big[\hat \rho _N \psi_ \beta (\vec y)^{\dagger}
\psi_ \alpha (\vec x)\big]=
\rho _{\alpha \beta }(\vec x ,\vec y)
\end{equation}
and
\begin{equation}
tr[\hat \rho_N \hat N]=\int \rho _{\alpha \alpha }(\vec x, \vec x) d\vec x
=N=\rho V
\end{equation}

We obtain the total nucleon energy as
\begin{eqnarray}
h_f & = & <f|tr[\hat \rho _N {\cal H}_N^{(0)}(\vec x)]|f> \nonumber \\
& = & {\gamma \over 6 \pi^2}k_f^3(M+{3\over 10}{k_f^2\over M})
\end{eqnarray}
where $\gamma =4$ for nuclear matter and $\rho $ and $k_f$ are related
by the equation $\rho =\gamma k_f^3/6\pi^2$.
With the meson field operator expansion as in Eq. (5) and (6)
we may write eq. (4) as
\begin{equation}
{\cal H}_M(\vec x)=a_i(\vec x)^\dagger \omega _x a_i(\vec x)
\end{equation}
Using Eq.(11) we now obtain kinetic energy density due
to the mesons as
\begin{eqnarray}
h_k & = &<f|{\cal H}_M (\vec x)|f> \nonumber \\
& = & 3 (2\pi)^{-3}\int d\vec k \omega(\vec k)sinh^2\tilde f(\vec k)
\end{eqnarray}
where $\omega (\vec k)=\sqrt{\vec k^2+m^2}$.
We next proceed to evaluate from Eq.(3) the interaction
energy density, with $\epsilon_x\simeq M$,
\begin{eqnarray}
h_{int} & = & <f|tr[\hat \rho _N {\cal H}_{int}(\vec x)]|f>\nonumber \\
&\simeq & {G^2\over 2 M}\rho<f|:\phi_i(\vec x)\phi_i(\vec x):|f>\label{eq:hi}
\end{eqnarray}
Using the Bogoliubov transformation (11) we have from equation ~(\ref{eq:hi})
\begin{equation}
h_{int}={G^2\rho\over 2 M}\cdot 3(2\pi)^{-3}\int {d\vec k
\over \omega(\vec k)} \left({sinh2\tilde f(\vec k)\over 2}
+sinh^2\tilde f(\vec k)\right)
\end{equation}
The meson energy density thus is given as
\begin{equation}
h_m=h_{int}+h_k\label{eq:intk}
\end{equation}
Now extremising equation ~(\ref{eq:intk}) with respect to $\tilde f(\vec k)$
we obtain the solution
\begin{equation}
tanh2\tilde f(\vec k)=-{G^2 \rho \over 2 M}\cdot {1\over {\omega ^2
(\vec k)+{G^2 \rho \over 2 M}}}\label{eq:tan1}
\end{equation}
We may compare the same with the results of Ref. [4] which corresponds
to the first term in the expansion of left hand side of Eq.
{}~(\ref{eq:tan1}). We then obtain the corresponding meson energy density
from the kinetic and interaction terms as
\begin{eqnarray}
h_m & = & h_k+h_{int}\nonumber \\
& = & -{3\over 2}\cdot (2\pi)^{-3}( {G^2 \rho \over 2 M} )^2\int{d\vec k\over
\omega (\vec k)[\omega (\vec k)(\omega (\vec k)^2+{G^2\rho\over M})^{1/2}
+(\omega (\vec k)^2+{G^2\rho\over 2M})]}\nonumber\\ & &
\end{eqnarray}
We note that Eq.(21) is not acceptable since the energy
density diverges. This happens because we have taken the pions
to be point like and assumed that they can approach as near each
other as they like, which is physically not correct. If we bring two
pions close to each other there will be an effective
force of repulsion because of their composite structure.
We thus assume a phenomenological term corresponding to
meson repulsion as
\begin{equation}
h_m^R=3a(2\pi)^{-3}\int (sinh^2\tilde f(\vec k))
e^{R_\pi^2k^2}d\vec k\label{eq:hmr1}
\end{equation}
where $a$ and $R_\pi$ are two parameters to be determined later.
Now extremising equation (16), (18) and (22) with respect to
$\tilde f (\vec k)$, we obtain
\begin{equation}
tanh2\tilde f(\vec k)=-{G^2 \rho \over 2 M}\cdot {1\over {\omega ^2
(\vec k)+{G^2 \rho \over 2 M}+a \omega (\vec k)e^{R_\pi^2k^2}}}.
\end{equation}
In place of Eq. (21) we now obtain the expression for $h_m$ as
\begin{eqnarray*}
h_m & = & h_k+h_{int}+h_m^R \\
& = &-{3\over 2}\cdot (2\pi)^{-3}\;( {G^2 \rho \over 2 M} )^2\cdot
\end{eqnarray*}
\begin{equation}
\int{d\vec k\over\omega^2}\cdot
\Big[{1\over{(\omega +a
e^{R_\pi^2k^2})^{1/2} (\omega +
a e^{R_\pi^2k^2}+{G^2 \rho\over M\omega})^{1/2}+(\omega +
a e^{R_\pi^2k^2)}+{G^2 \rho \over 2 M\omega }}}\Big]
\end{equation}
where $\omega =\omega (\vec k)$. Finally we have to include
the energy of repulsion which may arise from
vector meson interaction and/or from finite size of the nuclei.
We shall here parametrize the effect of such a
repulsion contribution by the simple form
\begin{equation}
h_R=\lambda \rho ^2
\end{equation}
where $\lambda $ is an arbitrary constant to be fixed from
phenomenology. We note that equation (25) can arise from a Hamiltonian
density given as
\begin{equation}
{\cal H}_R(\vec x)= \psi(\vec x)^\dagger\psi(\vec x)\int v_R(\vec x -\vec y)
\psi(\vec y)^\dagger\psi(\vec y)d\vec y
\end{equation}
where, when density is constant, we infact have
\[ \lambda =\int v_R(\vec r)d\vec r \]
We next minimise the energy per nucleon given by
\begin{equation}
E={(h_m+h_f+h_R)\over \rho }
\end{equation}
as a function of $\rho $, with the parameters $a$, $R_\pi$ and $\lambda $
to be subsequently fixed. Extremizing the single particle energy,
for $\lambda =0.54$ fm$^2$, $R_\pi=1.18$ fm and
$a=0.12$ GeV we have the single
particle energy given as $E=-15.03$ MeV at the saturation
density $\rho =0.19$ fm$^{-3}$corresponding to $k_f=1.42$ fm$^{-1}$.
The equation for E vs. $\rho $ is shown in Fig. 1 . The
incompressibility of the nuclear matter is given as
\begin{equation}
K=k_f^2 {\partial^2 E\over \partial k_f^2}=151.2 MeV
\end{equation}
 We may note that
the parameters and the results of the present analysis
do not differ significantly from those
of Ref. [4]. As may be seen in Eq.(20) or (23) Ref. [4] is an approximation
of the present framework.

We may also calculate the average pion number per nucleon $\rho _m$
given as
\begin{eqnarray}
\rho _m & = &<f|a_i^\dagger( \vec z) a_i(\vec z)|f> \nonumber \\
& = & 3 (2\pi)^{-3}\int d\vec k\; sinh^2\tilde f(\vec k)
\end{eqnarray}
which we plot in Fig. 4 as a function of nucleon density. This may be
relevant for heavy ion collisions where nuclear matter gets
compressed.

\section {Finite temperature formalism}

We shall now generalize the formalism developed in Sec. II for
finite temperatures. For this
purpose we shall use the methodology of thermofield dynamics.$^6$
In Appendix A, we summarize the salient features of the
same. Here Eq.(7) for the background pion will become modified
with introduction of extra thermal modes as illustrated in the
Appendix. In fact, we shall have the temperature dependent
background off-shell pion pair configuration given as
\begin{equation}
|f, \beta>=U_I(\beta)|f>=U_I(\beta)U|vac>
\end{equation}
where $U$ is the same as in Eq.(9), and $U_I(\beta )$ describes the change with
temperature.
The expression for this in terms of ordinary and thermal modes
is given as
\begin{equation}
U_I(\beta)=exp(B_I^\dagger(\beta )-B_I(\beta ))
\end{equation}
where, as in Eq.(A4) of Appendix A,
\begin{equation}
B_I(\beta )^\dagger={1 \over 2}\int \theta _B(\vec k, \beta )
b_i(\vec k)^\dagger\tilde b_i(-\vec k)^\dagger d\vec k
\end{equation}
In the above,
\begin{equation}
b_i(\vec k)^\dagger=Ua_i(\vec k)U^\dagger
\end{equation}
so that $b_i(\vec k)|f>=0$. Thus $b_i(\vec k)^\dagger$ creates excitations
over the zero temperature configuration given by $|f>$. As before
we shall calculate the energy expectation values, but here shall
minimise the thermodynamic potential as appropriate at finite
temperatures. To do so we note that a little algebra yields
\begin{eqnarray}
U^\dagger a_i(\vec k)U& = &(cosh\tilde f(\vec k))(cosh\theta_B(\vec k,\beta))
b_i(\vec k,\beta)+
(sinh\tilde f(\vec k))(cosh\theta_B(\vec k,\beta))
b_i(-\vec k,\beta)^\dagger \nonumber\\
&+&(sinh\tilde f(\vec k))(sinh\theta_B(\vec k,\beta))\tilde b_i(-\vec k,\beta)+
(cosh\tilde f(\vec k))(sinh\theta_B(\vec k,\beta))
\tilde b_i(\vec k,\beta)^\dagger
\nonumber \\
\end{eqnarray}
which again is a Bogoliubov transformation, and will be used for obtaining
the energy expectation values. The parallel unitary transformation as in
Eq.(30) for the temperature dependance in nucleon sector for fermions
is given as
\begin{equation}
U_{II}(\beta)=exp(B_{II}^\dagger(\beta )-B_{II}(\beta ))
\end{equation}
with
\begin{equation}
B_{II}(\beta )^\dagger={1\over 2}\int \theta _F(\vec k, \beta )
\psi_I(\vec k)^\dagger\tilde \psi_I(-\vec k)^\dagger d\vec k
\end{equation}
where $\theta _F(\vec k, \beta )$ will be determined later. As earlier
for thermal averages, we shall replace $\hat \rho _N \hat N$ in Eq. (13) by
\[ \hat \rho _N \psi_ \alpha (\vec x)^\dagger \psi_ \alpha (\vec x)
\rightarrow U_{II}(\beta )^\dagger
\psi_ \alpha (\vec x)^\dagger \psi_ \alpha (\vec x)U_{II}(\beta ).\]
We then have the nuclear matter density
\begin{eqnarray}
\rho (\beta )&=&<vac|U_{II}(\beta )^\dagger
\psi_ \alpha (\vec x)^\dagger \psi_ \alpha (\vec x)U_{II}
(\beta )|vac>\nonumber\\
& =&\gamma (2\pi)^{-3}\int d \vec k sin^2 \theta _F
\end{eqnarray}
As noted in the Appendix A, $sin^2 \theta _F$ is the distribution
function for the fermions. Clearly,
with $sin^2 \theta _F= \Theta(k_f-k)$ Eq. (37) gives
$\rho =\gamma k_f^3/6\pi^2$ of
zero temperature. $sin^2 \theta _F$ for the interacting system
will be determined here from the construction of the
thermodynamic potential.

We shall now calculate different contributions to the energy expectation
values corresponding to the Hamiltonian as in Eq. (1) and (4)
as well as Eq. (25). We thus have for the nucleon kinetic term
\begin{eqnarray}
h_f(\beta )&=& <vac|U_{II}(\beta )^\dagger \psi_I(\vec x)^\dagger
{(-\bigtriangledown_x^2)\over 2M}\psi_I(\vec x)U_{II}(\beta )|vac> \nonumber \\
&=&\gamma (2\pi)^{-3}\int d\vec k {k^2\over 2M}sin^2 \theta _F
\end{eqnarray}
The kinetic energy due to the mesons is given by
\begin{eqnarray}
h_k(\beta ) & = &<f,\beta|{\cal H}_M (\vec x)|f,\beta> \nonumber \\
& = & 3 (2\pi)^{-3}\int d\vec k \omega(\vec k)[sinh^2\tilde f(\vec k)
cosh2\theta_B(\vec k,\beta)+sinh^2\theta_B(\vec k,\beta)]
\label{eq:hk1}
\end{eqnarray}
where $\omega (\vec k)=\sqrt{\vec k^2+m^2}$.
In the above, we may note that when $\theta _B \rightarrow 0$ it
reduces to the Eq. (16).
We next derive the interaction energy density from Eq. (3) as
\begin{eqnarray}
h_{int}(\beta )&=& <f, \beta |U_{II}(\beta )^\dagger \psi_I(\vec x)^\dagger
\psi_I(\vec x){G^2\over 2 \epsilon _x}
\phi ^2(\vec x)U_{II}(\beta )|f, \beta > \nonumber \\
&\simeq & {G^2\over 2 M}\rho(\beta)<f,\beta|:\phi_i(\vec x)\phi_i(\vec x):
|f,\beta>\nonumber\\
&=& {G^2\rho(\beta )\over 2 M} I_{2M}
\label{eq:hint1}
\end{eqnarray}
where
\begin{equation}
I_{2M}={3\over(2\pi)^3}\int {d\vec k
\over \omega(\vec k)} \left({sinh2\tilde f(\vec k)cosh2\theta_B \over 2}
+sinh^2\tilde f(\vec k)cosh2 \theta _B+sinh^2 \theta _B\right).
\end{equation}
As in Eq. ~(\ref{eq:hmr1}) we shall now assume a phenomenological
term corresponding to meson repulsion due to composite structure of mesons
given as
\begin{equation}
h_m^R(\beta )=3a(2\pi)^{-3}\int \left(sinh^2\tilde f(\vec k)cosh2 \theta_B
+sinh^2 \theta _B\right) e^{R_\pi^2k^2} d\vec k\label{eq:hmr2}
\end{equation}
Finally, the nucleon repulsion term parallel to Eq.(25) is
\begin{equation}
h_R=\lambda \rho^{2}(\beta)
\end{equation}
where $\rho(\beta)$ is as given in Eq. (37). Thus the energy density
is given by
\begin{equation}
E(\beta)=(h_f(\beta)+h_m(\beta)+h_R(\beta))/\rho (\beta )
\end{equation}
where
\[h_m(\beta)=h_k(\beta)+h_m^R(\beta)+h_{int}(\beta)\]
as before.

The thermodynamic potential density $\Omega$ is given by
\begin{equation}
\Omega(\beta)=E(\beta)\rho -{\sigma\over \beta}-\mu\rho
\end{equation}
where the last term corresponds to nucleon number conservation with
$\mu $ as the chemical potential. We
may note that we shall be considering temperatures much below the nucleon
mass so that in the expression for $\rho(\beta)$ we do not include
antiparticle channel. The entropy density above is
$\sigma=\sigma_F+\sigma_B$ with $\sigma_F$ being the entropy in
fermion sector  given as$^6$
\[ \sigma_F=-{\gamma \over (2\pi)^3}\int d\vec k \Big[
sin^2 \theta _F(\vec k,\beta )
ln(sin^2 \theta _F(\vec k,\beta ))+cos^2 \theta _F(\vec k,\beta )
ln(cos^2 \theta _F(\vec k,\beta ))\Big]. \]
and simillarly the meson sector contribution $\sigma_B$ is given as
\[ \sigma_B=-{3\over (2\pi)^3}\int d\vec k \Big[ sinh^2 \theta _B(\vec k,\beta
)
ln(sinh^2 \theta _B(\vec k,\beta ))-cosh^2 \theta _B(\vec k,\beta )
ln(cosh^2 \theta _B(\vec k,\beta ))\Big]. \]
Thus the thermodynamic potential density now is a functional of
$\theta_F(\vec k, \beta)$, $\theta_B(\vec k, \beta)$ as well as the
pion dressing function $\tilde f(\vec k)$ which will of course
depend upon temperature. Extremisation of Eq (45) with respect to
$\tilde f(\vec k)$ yields
\begin{equation}
tanh2\tilde f(\vec k)=-{G^2 \rho \over 2 M}\cdot {1\over {\omega ^2
(\vec k)+{G^2 \rho \over 2 M}+a \omega (\vec k)e^{R_\pi^2k^2}}}
\end{equation}
which is of the same form as Eq. (23) for zero temperature case except that
now $\rho $ is temperature dependent through Eq. (37).
Similarly minimising the thermodynamic
potential with respect to $\theta _B(\vec k, \beta )$ we get
\begin{equation}
sinh^2 \theta _B={1\over e^{\beta \omega ^\prime}-1}
\end{equation}
where
\begin{equation}
\omega ^\prime=(\omega +{G^2 \rho \over 2M \omega }+a e^{R_\pi^2k^2})
cosh2\tilde f(\vec k)+{G^2 \rho \over 2 M \omega}sinh2\tilde f(\vec k)
\end{equation}
Once we substitute the optimised dressing as in Eq. (46), the above
simplifies to
\begin{equation}
\omega ^\prime=(\omega +{G^2 \rho \over M \omega }+a e^{R_\pi^2k^2})^{1/2}
(\omega +a e^{R_\pi^2k^2})^{1/2}
\end{equation}
which is different from $\omega $ due to interactions.
Further, minimising the thermodynamic potential with respect to
$\theta _F(\vec k, \beta )$ we have the solution
\begin{equation}
sin^2 \theta _F={1\over e^{\beta (\epsilon_F-\mu)}+1}
\end{equation}
with
\begin{equation}
\epsilon_F={G^2\over 2M}I_{2M}+2 \lambda \rho +{k^2\over 2M}
\end{equation}
where $I_{2M}$ is given in equation (41). We may note that the change
in $\epsilon_F$ above from $k^2/2M$ is also due to interaction.

\section {Results}

To calculate different thermodynamic quantities as functions of
baryon number density we first use Eq. (37) to calculate the chemical potential
$\mu$ in a self consistent manner with $\rho$ and $\mu$ occuring also inside
the
integrals through $sin^2\theta_F$ as in Eq. (50). Thus for each $\rho$,
we determine $\mu$ so that Eq. (37) is satisfied.
The ansatz functions $\theta _B$, $\theta _F$ and $\tilde f(\vec k)$
get determined analytically through the extremisation of the thermodynamic
potential and the parameters $a$, $\lambda $ and $R_\pi$ are as
obtained in Sec. II.

With the thermodynamic potential determined
as above, we calculate different thermodynamic
quantities. We first plot the binding energy per nucleon as a
function of baryon number density in Fig. 1 for temperatures zero
MeV, 5 MeV, 10 MeV and  15 MeV. As the temperature increases the
minimum shifts towards higher densities. This may be understood from
the fact that to compensate for the larger kinetic energy a large
value of $\rho$ is needed to give the minimum until the nuclear
repulsion effects take over and hence again energy increases. We may
compare our curves with those of Ref [2]. Our curves compared to
theirs are rather shallow. This is consistent with the fact that we
get a smaller value of compressibility ($\sim$ 150 MeV).

We next plot pressure as may be defined from thermodynamics$^7$
\begin{equation}
 P(\beta)=-\Omega (\beta)
\end{equation}
The same is plotted in Fig. 2. Vanishing of the pocket in the above curves
for temperatures around 15 MeV may be noted which could be an effect of
phase transition.

Entropy density is plotted as a function of nucleon density in Fig. 3
at temperatures 50 MeV, 100 MeV, 150 MeV and 200 MeV. The general
tendency appears to be quite similar to that of Ref. [2].

Finally, in Fig. 4 we plot the average number of pions per nucleon
$(\rho _m)$ as a function of nucleon density $\rho $. It is
interesting to note that this result is practically
independent of temperature. Therefore only the zero temperature graph
is plotted in Fig. 4. For low nucleon density $\rho _m$ rises
rather sharply and almost levels off at $\rho \geq 0.3$ fm$^{-3}$ $\simeq
\rho _s \times 1.5$ where $\rho _s$ is the density of stable
nuclear matter at zero temperature. We note that the rise
in meson number with compression may be
observable in heavy ion collisions.

\section{Discussions}

In the present model we have considered nuclear matter at zero temperature
and finite temperatures in a nonperturbative manner for the interacting
pion nucleon system using thermofield method.
This yields results similar to those of mean field approximation, with
$\sigma $-mesons {\it not} being needed, and, the calculations being
{\it quantum mechanical}.

Besides the aesthetic appeal, we note that it may have experimental
consequences. For example, the off shell $\pi^+ \pi^-$ pair as present
through pion dressing may
annihilate to hard photons with a probability in excess of what one may
expect otherwise. This is likely to depend on individual nuclei
which we may dress with pions as here.

\acknowledgments

The authors are thankful to A. Mishra, S.N. Nayak and Snigdha Mishra for
many discussions. SPM acknowledges to the Department of Science and
Technology, Government of India for the research grant
SP/S2/K-45/89 for financial assistance.

\appendix{}

We here summarize briefly the salient features of thermofield
dynamics$^6$ as used in the present paper.

In statistical mechanics, the thermal average of an operator
$\hat O$ is given as, with $\beta=1/kT$,
\begin{equation}
<{\hat O}>_{\beta }={Tr ({e^{-\beta H}}{\hat O})\over
{Tr (e^{-\beta H})}}
\end{equation}
where, the trace is taken over a complete
basis of states. First we note that in the zero temperature
limit, the above reduces to ground state expectation
value for the operator $\hat O$. This is easily seen as
\begin{eqnarray}
\lim_{\beta \rightarrow\infty}<\hat O>_{\beta }
&=&\lim_{\beta \rightarrow\infty}
{{<0\mid {\hat O}\mid 0>e^{-\beta \epsilon_{0}}
+<1\mid \hat O \mid 1>e^{-\beta  \epsilon _{1}}+\cdots}\over
{e^{-\beta \epsilon _ {0}}+e^{-\beta \epsilon _{1}}+\cdots}}\nonumber \\
&=&\lim_{\beta \rightarrow\infty}
{{<0\mid {\hat O}\mid 0>
+<1\mid \hat O \mid 1>e^{-\beta (\epsilon _{1}-\epsilon _{0})}+\cdots}\over
{1+e^{-\beta (\epsilon _{1}-\epsilon _{0})}+\cdots}}\nonumber\\
&=&<0\mid {\hat O}\mid 0>,
\end{eqnarray}
where $\mid 0>$ corresponds to the state with the lowest energy.
In thermofield method, one essentially generalises $(A2)$ to the
case of finite temperature and  defines a "thermal vacuum"
such that the statistical average reduces to an expectation value
with respect to the thermal vacuum.  Thus we want that
 for some $\mid 0( \beta )>$
the relationship
\begin{equation}
<{\hat O}>_{\beta }={Tr ({e^{-\beta H}}{\hat O})\over
{Tr (e^{-\beta H})}}\equiv <0(\beta) \mid {\hat O}\mid 0(\beta)>,
\end{equation}
where $\mid 0(\beta  )>$ is defined as the "thermal vacuum".
This can be done if
one doubles the degrees of freedom i.e. corresponding to every
physical operator $a$, a"tilde" operator $\tilde a$ is introduced.
In the above, ${\tilde a}(\vec k)$ are the new operators
named as "thermal modes". They are associated with negative
energy, with conventional quantisation, but, do not have any
physical significance in the sense of observation of these modes.
In zero temperature
vacuum, these modes are absent, so that conventional field theory
holds. At finite temperature the ground state is replaced by
$\mid 0(\beta)>$ given as
\begin{eqnarray}
| 0(\beta )>&\equiv& U_B(\beta  )| vac>\nonumber\\
&=&exp(\int \theta_B ({\vec k},\beta )(a(\vec k)^{\dagger }
{\tilde a}(-\vec k)^{\dagger} -h.c.)d{\vec k})\mid vac>,
\end{eqnarray}
where ${\tilde a}(-\vec k)^{\dagger}$ in the above
corresponds to the extra Hilbert space. It is now convenient to
define a thermal basis
\begin{equation}
\left(\begin{array}{c}a(\vec k,\beta )\\ \tilde a(-\vec k,\beta )^\dagger
\end{array}\right)
=U(\beta  )
\left(\begin{array}{c}a(\vec k) \\ \tilde a(-\vec k)^\dagger\end{array}\right)
U(\beta)^{-1},
\end{equation}
which amounts to the Bogoliubov transformation
\begin{equation}
\left(\begin{array}{c}a(\vec k,\beta ) \\ \tilde a(-\vec k,\beta )^\dagger
\end{array}\right)
=\left(\begin{array}{ll}\cosh \theta_B(\vec k,\beta)
& -\sinh \theta_B(\vec k,\beta)
\\ -\sinh \theta_B (\vec k,\beta) & \cosh \theta_B (\vec k,\beta)
\end{array}\right)\left(\begin{array}{c}
a(\vec k) \\ \tilde a(-\vec k)^\dagger\end{array}\right).
\end{equation}
$a(\vec k,\beta)$ and ${\tilde a}^\dagger(\vec k,\beta)$
are the annihilation and creation operators at temperature
$\beta={1\over KT}$ corresponding to
the thermal vacuum  such that
\[a(\vec k,\beta)\mid 0(\beta)>
=0={\tilde a}^\dagger(\vec k,\beta)| 0(\beta)>.\]
The function $\theta_B(k,\beta)$ is calculated through
minimization of thermodynamic potential density given as
\begin{equation}
\Omega=(E(\beta ) -{1\over \beta } \sigma_B +\mu N)
\end{equation}
where $\mu$, the chemical potential corresponds to a conserved
number, and the entropy density
\[ \sigma_B=-{3\over (2\pi)^3}\int d\vec k \Big[ sinh^2 \theta _B(\vec k,\beta
)
ln(sinh^2 \theta _B(\vec k,\beta ))-cosh^2 \theta _B(\vec k,\beta )
ln(cosh^2 \theta _B(\vec k,\beta ))\Big]. \]
For zero chemical potential and for free fields, extremization of the
free energy then yields
\begin{equation}
sinh^2 \theta _B={1\over e^{\beta \omega (\vec k, \beta )}-1}
\end{equation}
where we have taken the Hamiltonian density as ${\cal H}^0
=\int a^\dagger (\vec z) \omega _z a(\vec z)d\vec z$ so that
\begin{eqnarray*}
E(\beta )&= &<vac,\beta |{\cal H}^0 |vac \beta >\\
&=& {1\over (2\pi)^3}\int sin h^2 \theta _B \omega (\vec k) d\vec k
\end{eqnarray*}
If we substitute this value the free energy density becomes
\[ {\cal F}=E(\beta ) -{1\over \beta }\sigma _B={1\over \beta }
{1\over (2\pi)^3}\int ln(1-e^{-\beta\omega (\vec k)})\; d \vec k\]
which is the same as derived through a more convenational
treatment of temperature dependent quantum field theory.$^8$
As stated, we do here the analysis through thermofield dynamics
since the calculations are simpler.
In case of interacting field however the solution for $\theta _B(\vec k,
\beta)$
will not be given by equation (A8) and will depend upon the interaction.

Similarly for the fermionic sector the thermal vacuum will be given by
\begin{eqnarray}
| 0(\beta )>&\equiv& U_F(\beta  )| vac>\nonumber\\
&=&exp(\int \theta_F ({\vec k},\beta )(\psi_I(\vec k)^{\dagger }
{\tilde \psi}_I(-\vec k)^{\dagger} -h.c.)d{\vec k})\mid vac>
\end{eqnarray}
where ${\tilde \psi}_I^\dagger $ corresponds to the creation of the
fermionic thermal modes. The entropy is given as
\[ \sigma_F=-{\gamma \over (2\pi)^3}\int d\vec k \Big[
sin^2 \theta _F(\vec k,\beta )
ln(sin^2 \theta _F(\vec k,\beta ))+cos^2 \theta _F(\vec k,\beta )
ln(cos^2 \theta _F(\vec k,\beta ))\Big]. \]

The thermal Bogoliubov transformation is now given as
\begin{equation}
\left(\begin{array}{c}\psi_I(\vec k,\beta ) \\
{\tilde \psi}_I(-\vec k,\beta )^\dagger
\end{array}\right)
=\left(\begin{array}{ll}\cos \theta_F(\vec k,\beta)
& -\sin \theta_F(\vec k,\beta)
\\ -\sin \theta_F (\vec k,\beta) & \cos \theta_F (\vec k,\beta)
\end{array}\right)\left(\begin{array}{c}
\psi_I(\vec k) \\ {\tilde \psi}_I(-\vec k)^\dagger\end{array}\right).
\end{equation}
Parallel to equation (A8) the function of $\theta _F(\vec k, \beta )$
is given as
\begin{equation}
sin^2 \theta _F= {1\over e^{\beta (\omega (\vec k, \beta )-\mu)}+1}
\end{equation}
where $\mu$ is the chemical potential corresponding to baryon
number conservation and $\omega (\vec k, \beta )=\sqrt{ k^2+M^2}$
for free fermions of mass $M$,
so that when statistics is known, the corresponding Bogoliubov
tranformation relating zero temperature ground state with the thermal
ground state of the extended Hilbert space is known.
The ground state or the "thermal vacuum" obviously contains particles
with appropriate distributions.

The methodology enables us to replace mixed states of statistical
mechanics by pure states in an extended Hilbert space while generating
correct distribution functions. The extra thermal modes enable us to
do this.

\figure{Binding energy per nucleon E as a function of nuclear
matter density $\rho $ at temperatures T=0 MeV, 5 MeV, 10 MeV and 15 MeV.
The gradual shift in the energy minimum towards higher densities may be
noted.}
\figure{Pressure P as a function of nuclear matter density $\rho $ at
temperatures T=0 MeV, 5 MeV, 10 MeV, 15 MeV and 20 MeV. The vanishing
of $`$pocket' at about 15 MeV may be taken as a signature of
phase transition.}
\figure{Entropy density $\sigma $ as a function of nuclear matter
density $\rho$ at temperatures T=50 MeV, 100 MeV, 150 MeV and 200 MeV.}
\figure{Average number of pions per nucleon $\rho _m$ as a function of
nuclear matter density $\rho $. Since the results are effectively
independent of temperature, only the zero temperature graph is
plotted.}

\begin{references}
\begin{enumerate}
\item  J.D. Walecka, Ann. Phys. {\bf 83} (1974) 491; B.D. Serot
and J.D. Walecka, Adv. Nucl. Phys. {\bf 16} (1986) 1; S. Gmuca,
J. Phys. \ G {\bf 17}, (1991) 1115.
\item  R.J. Furnstahl and B.D. Serot, Phys. \ Rev. \ C {\bf 41}
(1990) 262.
\item  Y.K. Gambhir and P. Ring, Phys. \ Lett. \ B {\bf 202}
(1988) 2
\item  A. Mishra, H. Mishra and S.P. Misra, Int. J. Mod. Phys. \ A
{\bf 7} (1990) 3391.
\item  J.M. Eisenberg, Phys. \ Lett \ B {\bf 104} (1988) 353;
H. Jung, F. Beck and G.A. Miller, Phys. \ Rev. \ Lett. {\bf 62} (1989) 2357.
\item  H. Umezawa, H. Matsumoto and M. Tachiki {\it Thermofield
dynamics and condensed states} (North Holand, Amsterdam, 1982)
\item  See for example in $``$ Quantum Theory of Many Particle
System" A.L. Fetter and J.D. Walecka, {\it McGraw Hill Book
Company}, 1971.
\item  I.Dolan and R. Jackiew, Phys. \ Rev. \ D {\bf 9},
(1974) 3320.
\end{enumerate}

\end{references}
\end{document}